\documentclass[iop]{emulateapj}

\usepackage{atbegshi}

\def\ltsima{$\; \buildrel < \over \sim \;$}
\def\simlt{\lower.5ex\hbox{\ltsima}}
\def\gtsima{$\; \buildrel > \over \sim \;$}
\def\simgt{\lower.5ex\hbox{\gtsima}}
\def\kms{{\rm\,km\,s^{-1}}}

\def\kpc{{\rm\,kpc}}

\def\msun{{\rm\,M_\odot}}
\def\lsun{{\rm\,L_\odot}}

\def\s{\ifmmode \widetilde \else \~\fi}
\def\={\overline}

\def\spose#1{\hbox to 0pt{#1\hss}}

\def\eg{{e.g.,\ }}
\def\ie{{ i.e.,\ }}
\def\lta{\mathrel{\spose{\lower 3pt\hbox{$\mathchar"218$}}
     \raise 2.0pt\hbox{$\mathchar"13C$}}}
\def\gta{\mathrel{\spose{\lower 3pt\hbox{$\mathchar"218$}}
     \raise 2.0pt\hbox{$\mathchar"13E$}}}
\def\Dt{\spose{\raise 1.5ex\hbox{\hskip3pt$\mathchar"201$}}}    % upper case
\def\dt{\spose{\raise 1.0ex\hbox{\hskip2pt$\mathchar"201$}}}    % lower case

\def\dotsfill{\leaders\hbox to 1em{\hss.\hss}\hfill}

\shorttitle{NGC1052-DF2}
\shortauthors{N. F. Martin et al.}

\begin{document}


\title{Current velocity data on dwarf galaxy NGC1052-DF2 do not constrain it to lack dark matter}


\author{Nicolas F. Martin$^{1,2}$, Michelle L. M. Collins$^3$, Nicolas Longeard$^1$, Erik Tollerud$^4$}

\email{nicolas.martin@astro.unistra.fr}

\altaffiltext{1}{Universit\'e de Strasbourg, CNRS, Observatoire astronomique de Strasbourg, UMR 7550, F-67000 Strasbourg, France}
\altaffiltext{2}{Max-Planck-Institut f\"ur Astronomie, K\"onigstuhl 17, D-69117 Heidelberg, Germany}
\altaffiltext{3}{Department of Physics, University of Surrey, Guildford, GU2 7XH, Surrey, UK}
\altaffiltext{4}{Space Telescope Science Institute, 3700 San Martin Dr., Baltimore, MD 21218, USA}

\begin{abstract}
It was recently proposed that the globular cluster system of the very low surface-brightness galaxy NGC1052-DF2 is dynamically very cold, leading to the conclusion that this dwarf galaxy has little or no dark matter. Here, we show that a robust statistical measure of the velocity dispersion of the tracer globular clusters implies a mundane velocity dispersion and a poorly constrained mass-to-light ratio. Models that include the possibility that some of the tracers are field contaminants do not yield a more constraining inference. We derive only a weak constraint on the mass-to-light ratio of the system within the half-light radius ($M/L_V<6.7$ at the 90-percent confidence level) or within the radius of the furthest tracer ($M/L_V<8.1$ at the 90-percent confidence level). This limit may imply a mass-to-light ratio on the low end for a dwarf galaxy but many Local Group dwarf galaxies fall well within this contraint. With this study, we emphasize the need to reliably account for measurement uncertainties and to stay as close as possible to the data when determining dynamical masses from very small data sets of tracers.
\end{abstract}

\keywords{galaxies: kinematics and dynamics --- methods: statistical}

\section{Introduction}
The dwarf galaxy NGC1052-DF2 is a satellite of the elliptical NGC 1052 ($M_V\simeq-19.4$) discovered by \citet{karachentsev00} and later studied in detail by the Dragonfly experiment \citep{vandokkum15b}. It is a very low surface brightness system, owing to its large half-light radius ($M_V\sim-15.3$; $r_\mathrm{half} \sim2.2\kpc$; $\mu_0=24.4\textrm{\,mag}/\textrm{arcsec}^2$; \citealt{vandokkum18a}). The presence of easily identified globular clusters in the system allowed \citet[][hereafter vD18b]{vandokkum18a} to explore the dynamics of this so-called ``ultra-diffuse galaxy''. From the velocities they obtained with LRIS and DEIMOS on the Keck telescopes, the authors isolate 10 likely member globular clusters (GCs), centered around $cz=1803 \kms$. vD18b show that an rms estimate of the velocity dispersion of this sample yields $\sigma_\mathrm{rms}\sim14.3\kms$, while the use of a biweight dispersion \citep{beers90} yields a smaller value $\sigma_\mathrm{rms}\sim8.4\kms$. This is expected since this latter technique, which they favor, removes potential outliers to the distribution and produces a colder dispersion. After accounting for these uncertainties and under the hypothesis that the furthermost point (GC98) is an outlier, vD18b estimate an intrinsic velocity dispersion of $\sigma_\mathrm{int}=3.2^{+5.5}_{-3.2}\kms$.

However, it is well known that for such small samples of tracers that also have velocity uncertainties of order the measured velocity dispersion, results are extremely sensitive to the technique used and to the way the uncertainties are handled. This is a state of affair that is, unfortunately, too common for the study of the dynamics of very faint dwarf galaxies in the Local Group for which samples are often restricted to 5--20 stars with velocities \citep[\eg][]{martin07a,simon07}. This community has converged on statistical methods that infer the velocity dispersion of a system by simply building a generative model for the data (\eg \citealt{hogg10}; to measure a velocity dispersion, we would use a single Gaussian distribution, or the sum of a Gaussian distribution with a simple contamination model that can handle outliers) and evaluating the posterior probability distribution. In favorable cases, the latter can potentially be summarised by its associated modes if it is well behaved.

In this letter, we revise the estimation of the velocity dispersion of NGC1052-DF2 by building such a generative model and sampling the posterior PDF of the intrinsic velocity dispersion. We show that the current data does not imply a vanishingly small velocity dispersion (and mass-to-light ratio) for NGC1052-DF2 and that, in fact, it is compatible with expectations from dynamically hot (\ie dark-matter dominated) Local Group dwarf galaxies.

\section{Method and results}
We base our analysis on the sample of 10 GC velocities presented in \citet{vandokkum18b} and vD18b, with their associated uncertainties.

%\begin{table}
%\caption{\label{data}Velocities of the 10 GCs of NGC1052-DF2 from vD18b and \citet{vandokkum18b}.}
%\begin{tabular}{lllrr}
%\# & RA & Dec & $cz (\kms)$ & $\Delta v (\kms)$\\
%\hline
%39 & 2:41:45.07 & $-8$:25:24.9 & $1818\pm7$ & $15\pm7$ \\
%59 & 2:41:48.08 & $-8$:24:57.5 & $1799^{+16}_{-15}$ & $-4^{+16}_{-15}$ \\
%71 & 2:41:45.13 & $-8$:24:23.0 & $1805^{+6}_{-8}$ & $2^{+6}_{-8}$ \\
%73 & 2:41:48.22 & $-8$:24:18.1 & $1814\pm3$ & $11\pm3$ \\
%77 & 2:41:46.54 & $-8$:24:14.0 & $1804\pm6$ & $1\pm6$ \\
%85 & 2:41:47.75 & $-8$:24:05.9 & $1801^{+5}_{-6}$ & $-2^{+5}_{-6}$ \\
%91 & 2:41:42.17 & $-8$:23:54.0 & $1802\pm10$ & $-1\pm10$ \\
%92 & 2:41:46.90 & $-8$:23:51.1 & $1789^{+6}_{-7}$ & $-14^{+6}_{-7}$ \\
%98 & 2:41:47.34 & $-8$:23:35.2 & $1764^{+11}_{-14}$ & $-39^{+11}_{-14}$ \\
%101 & 2:41:45.21 & $-8$:23:28.3 & $1800^{+13}_{-14}$ & $-3^{+13}_{-14}$ \\
%\end{tabular}
%\end{table}

\subsection{Model with no contamination}
\label{no_cont}

We first assume that all 10 GCs are members of NGC1052-DF2, with velocities $v_i$ and velocity uncertainties $\delta_{v,i}$. In this case, our generative model is a simple Gaussian function with mean $\langle v\rangle$ and intrinsic dispersion, $\sigma_\mathrm{int}$. The likelihood function can be expressed as

\begin{eqnarray}
\mathcal{L} = \prod_{i=1}^{i\leq10}\frac{1}{\sqrt{2\pi}\sigma_\mathrm{obs}} \exp\Big(-0.5\Big(\frac{v_i-\langle v\rangle}{\sigma_\mathrm{obs}}\Big)^2\Big),\\
\textrm{with } \sigma_\mathrm{obs}^2 = \sigma_\mathrm{int}^2+\delta_{v,i}^2.
\end{eqnarray}

\noindent Since the uncertainties $\delta_{v,i}$ provided by vD18b are asymmetric, we use the positive uncertainty when $v_i<\langle v\rangle$ and the negative one otherwise.

\begin{figure}
\begin{center}
\includegraphics[width=0.75\hsize,angle=270]{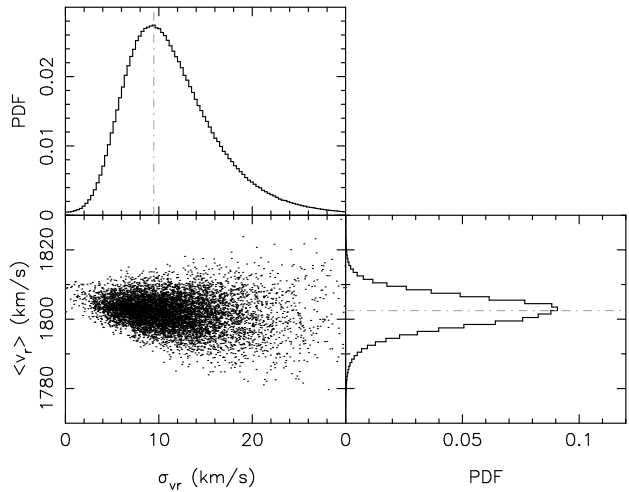}
\caption{\label{without_cont} Joint PDF of the two-parameter Gaussian model (bottom-left) and the marginalized PDF for the mean velocity $\langle v_r\rangle$ (right) and the velocity dispersion $\sigma_\mathrm{int}$ (top). This model yields $\sigma_\mathrm{int}= 9.5^{+4.8}_{-3.9}\kms$.}
\end{center}
\end{figure}

We assume uniform priors on $\langle v\rangle$ and $\sigma_\mathrm{int}$ over the ranges 1750 to $1850\kms$ and 0 to $30\kms$, respectively. We then sample the posterior PDF with our own Markov Chain Monte Carlo algorithm \citep{martin16a,longeard18}. The resulting joint PDF is shown in Figure~\ref{without_cont}, along with the marginalized PDFs for the two parameters. The PDF on the intrinsic velocity dispersion of NGC1052-DF2 is well behaved and yields a significantly higher dispersion than the one reported by vD18b: $\sigma_\mathrm{int}= 9.5^{+4.8}_{-3.9}\kms$ ($<18.8\kms$ at the 90-percent confidence level) vs. $\sigma_\mathrm{int}=3.2^{+5.5}_{-3.2}\kms$ ($<10.5\kms$ at the 90-percent confidence level). Note that our measurement is by-design corrected for the velocity uncertainties as those are specifically included in the model. Our inference is compatible with the rms estimate of vD18b ($\sigma_\mathrm{rms}\sim12.2\kms$); this is expected since of all three methods used by vD18b, the rms estimate most closely resembles our formalism.

\subsection{Priors}
The inference described above assumes a uniform prior on $\sigma_\mathrm{int}$ but it is known that such a prior can be biased for small values. We also test the use of Jeffreys's prior, which does not suffer from this bias, but has the uncomfortable property of being improperly defined (i.e. the PDF does not integrate to unity) if it is not bound at the lower end. Doing so and forcing $\sigma_\mathrm{int}>1\kms$ yields $\sigma_\mathrm{int}=7.4^{+4.5}_{-3.3}\kms$ ($<15.5\kms$ at the 90-percent confidence level), which does not significantly change our inference. Alternatively, one can argue that, since the dynamical mass of NGC1052-DF2 is the physical quantity we aim to constrain and since this quantity scales as $\sigma_\mathrm{int}^2$, it would be more appropriate to assume a uniform prior on $\sigma_\mathrm{int}^2$. Unsurprisingly, doing so yields larger value for the most likely intrinsic dispersion, with $\sigma_\mathrm{int}=13.1^{+6.6}_{-4.5}\kms$ ($<27.2\kms$ at the 90-percent confidence level).

While changing the prior on $\sigma_\mathrm{int}$ does not change the main conclusion of this paper (the velocity dispersion of the NGC1052-DF2 velocity sample is not very well constrained), the fluctuations on the constraint stemming from the choice of prior displays the poor constraining power of the data set.

\subsection{Model with contamination}
\begin{figure}
\begin{center}
\includegraphics[width=0.75\hsize,angle=270]{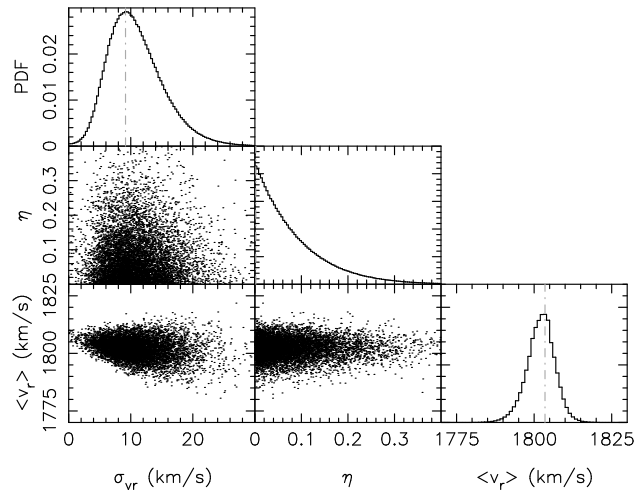}
\caption{\label{with_cont} Joint PDF for the three-parameter model with a Gaussian component and an uniform contamination population. The marginalized PDF for the mean velocity $\langle v_r\rangle$, the fraction of contaminants $\eta$, and the velocity dispersion $\sigma_\mathrm{int}$ are also shown. Despite the contamination component, the PDF on the velocity dispersion remains similar to that of Figure~\ref{without_cont}, with $\sigma_\mathrm{int}= 9.2^{+4.8}_{-3.6}\kms$.
}
\end{center}
\end{figure}

A Kolmogorov-Smirnov test yields a high probability of 0.4 to 0.8 that the sparse data set is drawn from the range of models constrained in sub-section~\ref{no_cont}. It is therefore not possible to reject the simple Gaussian model as a bad model for this data set. Nevertheless, it is a priori possible that the sample of 10 GCs includes some contamination by field GCs (\eg from the neighboring NGC~1052) and we now test a model that allows for contamination. We assume a uniform contamination model, $\mathcal{U}$ over the range $1750<v_r<1850\kms$. With $\eta$ the fraction of the data that is in the contamination, the likelihood function becomes 

\begin{equation}
\mathcal{L} = \prod_{i=1}^{i\leq10}\Bigg[\eta\,\mathcal{U}+\frac{1-\eta}{\sqrt{2\pi}\sigma_\mathrm{obs}} \exp\Big(-0.5\Big(\frac{v_i-\langle v\rangle}{\sigma_\mathrm{obs}}\Big)^2\Big)\Bigg],
\end{equation}

\noindent with $\sigma_\mathrm{obs}$ as defined in equation (1). The resulting PDFs are shown in Figure~\ref{with_cont} for uniform priors. Interestingly, the inference on the intrinsic velocity dispersion of the GC sample remains unchanged, despite $\eta$ reaching an upper limit of $\sim0.3$. While this may seem surprising at first, it can easily be explained by the datum with the most discrepant velocity (GC98; $v = 1764^{+11}_{-14}\kms$) having one of the largest velocity uncertainties. The model does not feel the need to separate this datum and fold it in the contamination model (indeed, that GC has the high probability of $\sim0.9$ to belong to the dwarf galaxy part of the model). After marginalization, we infer $\sigma_\mathrm{int}=9.2^{+4.8}_{-3.6}\kms$ ($<17.3\kms$ at the 90-percent confidence level) for our baseline model with contamination. If we use a less constraining contaminant model using a second Gaussian with only loose priors on the contamination (uniform from 1700 to $1900\kms$ for the center and uniform between 100 to $200\kms$ for the dispersion of this Gaussian representing the contamination), we get $\sigma_\mathrm{int} = 11.4^{+5.8}_{-4.5}\kms$. Finally, even if we nevertheless decide to forego the outcome of the modeling with contamination and abruptly remove GC98 from the data set (which is not advisable) to fit a single Gaussian-model to the velocities of the remaining 9 GCs, we still only infer $\sigma = 7.1^{+3.6}_{-3.0}\kms$ ($<14.3\kms$ at the 90-percent confidence level). These variable results for different contamination assumptions highlight the challenges in interpreting such small-number datasets, while also demonstrating that these cases yield dispersions significantly higher than the vD18b limit.

In the following we will use the model with the uniform contamination as our baseline model since it is among the most agnostic models discussed above.

\subsection{Impact on the mass-to-light ratio}
\begin{figure}
\begin{center}
\includegraphics[width=0.75\hsize,angle=270]{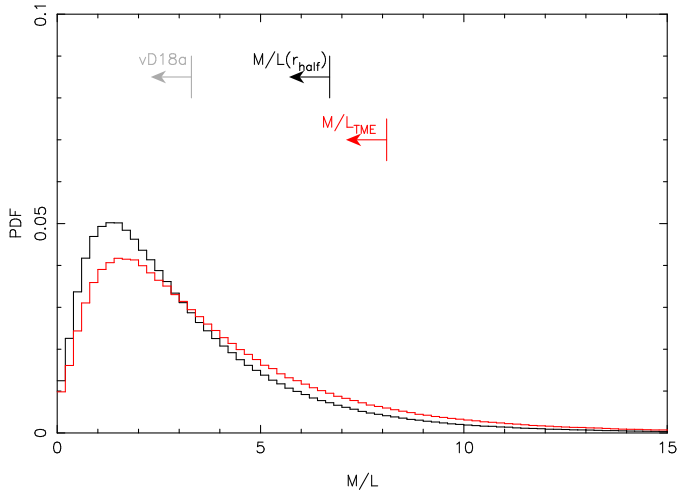}
\caption{\label{ML} Mass-to-light ratio of NGC1052-DF2 infered from the marginalized velocity dispersion PDF of the model with contamination and for the \citet{walker09} mass estimator within the half-light radius (black) and the TME mass estimator of \citet{watkins10} favored by vD18b (red). In both cases, we infer a much less strict limit, as can be seen by the 90-percent confidence limits implied by our analysis (black and red limits and arrow) and that of vD18b (gray limit and arrow).
}
\end{center}
\end{figure}

To infer the mass-to-light ratio of NGC1052-DF2, we rely on the velocity dispersion from the model with contamination and use the mass estimator of \citet{walker09} that provides the mass within the half-light radius of the dwarf galaxy ($\sim2.2\kpc$) under the usual assumption of dynamical equilibrium and sphericity. This estimator yields $M(r_\mathrm{half})<3.7\times10^8\msun$ at the 90-percent confidence level. Since this radius naturally includes half of the light of the system ($\sim0.55\times10^8\lsun$), we can infer the mass-to-light ratio $M/L(r_\mathrm{half},V)$ within the half-light radius. The corresponding PDF is shown in black in Figure~\ref{ML} and yields an upper limit of 6.7 at the 90-percent confidence level. \citet{wolf10} describe an alternate mass estimator to the one of \citet{walker09} that, beyond highlighting the difficulty of modeling the dynamical mass from a population of tracers, yields larger masses than the ones we give here ($M/L(r_\mathrm{half},V)<10.7$). The difference is driven by different choices for the profiles of the tracers, their (axi)symmetry and/or anisotropy assumption (see the discussion in appendix C of \citealt{wolf10}). We focus on the \citet{walker09} estimator to allow for an easier comparison with vD18b but recognize that the mass-to-light ratio limit of NGC1052-DF2 would be even higher than the one we infer if we had used the \citet{wolf10} estimator.

The mass estimator favored by vD18b and based on \citet{watkins10} gives the mass within the last datum, i.e. within $7.6\kpc$ for the sample of GCs\footnote{Based on the structural parameters of NGC1052-DF2 (vD18b), this radius includes 98 percent of the overall luminosity of the dwarf galaxy, or $\sim1.08\times10^8\lsun$.}. The resulting mass-to-light ratio inference is similar but slightly less constrained (the red curve in Figure~\ref{ML}; $M/L_\mathrm{TME}<8.1$ at the 90-percent confidence level).

Both mass estimators are therefore consistent with each other and the data set is not strongly constraining, contrary to the finding of vD18b who found $M/L_{\mathrm{TME},V}<3.3$ at the 90-percent confidence level. It is also worth noting that folding in the uncertainties on $r_\mathrm{half}$ and $L_V$ would make the constraint weaker but vD18b unfortunately do not provide those for their updated measurement of the size and luminosity of NGC1052-DF2. As such, the confidence limits provided here should only be seen as lower limits.

\subsection{Additional tests}
\subsubsection{Measuring the velocity dispersion by resampling the observed data}

\begin{figure}
	\includegraphics[width=\columnwidth]{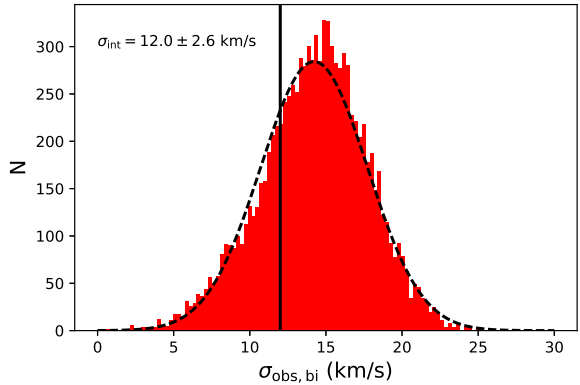}
    \caption{Results for measuring the observed biweight-midvairance dispersion from 10,000 resamples of the vD18b dataset. Here, the original velocities are perturbed within their $1\sigma$ uncertainties as described in the text. The mean observed bi-weight for the sample comes out as $\sigma_\mathrm{obs,bi}=14.3\pm3.5\kms$, giving $\sigma_\mathrm{int,bi}=12.0\pm2.5\kms$, higher than the 90\% upper limit from vD18b, and consistent with our MCMC analysis.}
    \label{fig:mcbw}
\end{figure}

Even though it amounts to make the data more noisy than they truly are and we do not recommend it, a common technique for measuring the dispersion from a small number of data points with significant uncertainties (i.e. similar to the size of the dispersion this is being measured), is to run a Monte Carlo resampling of the data. Here, we take the observed velocities of the vD18b sample, and perturb them based on their uncertainties by randomly sampling from a Gaussian centered on the velocity measurement, with a dispersion equal to the uncertainties quoted by vD18b. We then follow their method for measuring the observed dispersion by recomputing the bi-weighted mid-variance for this perturbed sample. We repeat this process 10,000 times, resulting in a distribution of values for $\sigma_{\rm obs, bi}$ (see Figure~\ref{fig:mcbw}). From this process, we can use the mean and standard deviation of the distribution as a value for the observed dispersion, giving $\sigma_\mathrm{ob,bi}=14.3\pm3.5\kms$ (very comparable to the observed r.m.s dispersion from vD18). Following this, we must also correct for the effects of the observational uncertainties, which will inflate this measurement. We follow the process of \citet{pryor93}, using the average uncertainty from our 10,000 realisations, resulting in $\sigma_\mathrm{int,bi}= 12.0\pm2.5\kms$. This value is considerably higher than the $\sigma_\mathrm{int}=3.2^{+5.5}_{-3.2}\kms$ from vD18b, and lies above their proposed upper limit of  $\sigma_\mathrm{int}<10.5\kms$, but is consistent with the value we compute with our generative model (with or without contamination).

\subsubsection{On the reliability of using small samples of globular clusters to compute the `true' dispersion}

Two key issues with interpreting any measured velocity dispersion in this instance are (1) the small number of tracers available and (2) knowing whether these are truly relaxed tracers of the underlying dark matter halo. For the latter, we know from observations of the outskirts of both the Milky Way and Andromeda that GCs are often associated with substructure at large radii (\eg \citealt{mackey10, veljanoski14}). In Andromeda in particular, between 50-80\% of all GCs at distances beyond $30\kpc$ show both spatial and kinematic correlations with stellar streams \citep{mackey10}, meaning that they are not fully relaxed mass tracers.

\begin{figure}
	\includegraphics[width=\columnwidth]{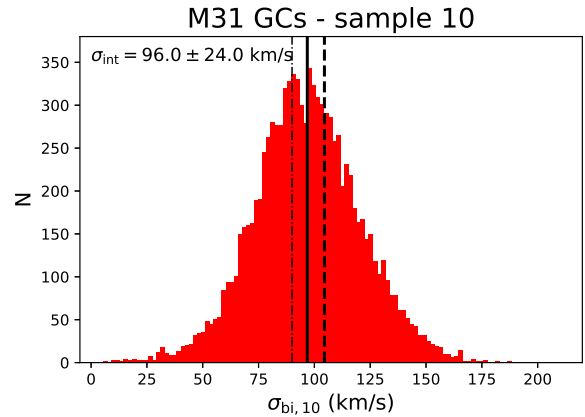}
    \caption{Results from randomly sampling 10 GCs from M31's outer cluster population, and measuring their dispersion from the biweight-midvariance, as in vD18. The mean of this analysis is shown as the solid line, while the value from the full sample of 74 clusters is shown as the dashed line. The value of the velocity dispersion from M31 halo stars is shown as the dash-dotted line.}
    \label{fig:m31}
\end{figure}

Given point (2), the effects of point (1) could be severe. Measuring a single dispersion from 10 tracers that may not be relaxed could lead to either a significant over- or under-estimate of the halo velocity dispersion. This can be straightforwardly demonstrated using the GC system of M31. We take the kinematics for 72 clusters from \citet{veljanoski14}. As the globular cluster system of M31 is known to rotate, we use their rotation-corrected velocities to ensure we are not artificially inflating our measured mass. We then randomly draw 10 clusters and measure the biweight-midvariance of their velocity distribution, following the technique used by vD18b. This sample has a much larger intrinsic dispersion than DF2, but the data are of similar quality (mean velocity uncertainties of $\sim10\kms$). Repeating this process 10,000 times gives us a distribution of observed velocity dispersions (see Figure~\ref{fig:m31}) that we can compare to both the biweight from the full sample ($\sigma_\mathrm{bi,all} = 105.0\kms$, dashed line, Figure~\ref{fig:m31}), and the average velocity dispersion of the M31 halo from its stars ($\sigma_\mathrm{M31_{stars}}\sim90\kms$, dash-dotted line; \citealt{gilbert18}).

The result here is clear: a single biweight dispersion measure from 10 GCs can give a huge range of dispersion measures. The mean value from this redraw gives $\sigma_{\rm bi, 10}=96\pm24\kms$, but the tails extend to far higher and smaller values. Such a large statistical uncertainty would mean that, from a sample of 10 GCs in M31, halo masses ranging from $0.2<M<1.2\times10^{12}M_\odot$ could be measured within the 90\% confidence limit. Given that M31 has a stellar mass of $\sim10^{11}M_\odot$ \citep{sick15,williams17}, the mass-to-light ratio could also be compatible with no dark matter based on this analysis.

\section{Discussion}
It is evident from Figure~\ref{ML} that the current velocity data set on NGC1052-DF2 is not very constraining beyond pointing out that the dwarf galaxy is not massively dominated by dark matter. At the moment, it is not possible to rule out any mass-to-light ratio below $M/L<6.7$ within the half-light radius or $M/L<8.1$ within the radius covered by the tracers (at the 90-percent confidence level in both cases). Could NGC1052-DF2 host no dark matter and its inferred mass (or mass-to-light ratio) be entirely consistent with an old stellar population ($M/L_V\sim2$)? Certainly, but so could a much more mundane, dark-matter dominated mass-to-light ratio.

The mass-to-light ratio of NGC1052-DF2 is compatible with that of other nearby dwarf galaxies. For instance, IC~1613 shares the luminosity of NGC1052-DF2, has a radius that is only half as small and a velocity dispersion of $10.8^{+1.0}_{-0.9}\kms$ from which \citet{kirby14} inferred $M/L_V(r_\mathrm{half})=2.2\pm0.5$. The M31 companions Cas~III and Lac~I, albeit somewhat fainter, share similar properties to those of NGC1052-DF2: their large half-light radii ($\sim1.5\kpc$; \citealt{martin13a}) and velocity dispersion $\sim10\kms$ imply mass-to-light ratios ($M/L_V(r_\mathrm{half})=8^{+9}_{-5}$ and $15^{+12}_{-9}$, respectively; \citealt{martin14b}) that are entirely compatible with the constraint on NGC1052-DF2\footnote{The fairly large uncertainties on $M/L_V(r_\mathrm{half})$ for these two systems, despite being based on 100--200 tracers further imply that the 10 NGC1052-DF2 tracers with velocities are unlikely to yield a strong constraint.}. The well-study Milky Way satellite dwarf galaxy Fornax also shares similar properties \citep{irwin95,walker09}. Finally, NGC1052-DF2's velocity dispersion and mass, despite being poorly constrained, fall perfectly on the \citet{walker09} universal mass profile proposed for Local Group dwarf galaxies. It also follows the locus of most dwarf galaxies in the $M/L$ vs. $M$ plane, contrary to the peculiar dwarf galaxy Dragonfly~44 that appears exceptionally massive \citep[][their Figure~3]{vandokkum16}. A conservative and cautious approach would therefore be to conclude that the mass-to-light ratio of NGC1052-DF2 appears to be the low end of that measured for other dwarf galaxies, but share the properties of other local dwarf galaxies and relies on a noisy measurement. Other ``ultra-diffuse dwarf galaxies'' studied with data sets of similar quality also yield only weak constraints on the dark-matter content \citep{toloba18}. Significant additional proof is required before claiming a lack of dark matter in NGC1052-DF2, even more so since rotation could also be present and its contribution to the dynamics of the galaxy could further increase its dynamical mass. An independent study by \citealt{laporte18} shows that NGC1052-DF2 can comfortably live in a dark matter halo of $10^9\msun$ or even $10^{10}\msun$ within the uncertainties.

The different conclusions reached by vD18b and this study show the difficulty in extracting information from a small velocity data set, especially when the measurement uncertainties on the individual data points are of order the dispersion that is being inferred. In such cases, reverting back to the simplest model and techniques (using a generative model) yields more robust and tractable results.

\acknowledgments

We thank the reviewer and the statistics editor for their celerity in evaluating this letter. We thank Annette Ferguson, Andrew Hearin, and Justin Read for fruitful discussions. N. F. Martin gratefully acknowledges the Kavli Institute for Theoretical Physics in Santa Barbara and the organizers of the ``Cold Dark Matter 2018'' program, during which some of this work was performed. This research was supported in part by the National Science Foundation under Grant No. NSF PHY11-25915. N. F. Martin and N. Longeard also acknowledge support by the Programme National Cosmology et Galaxies (PNCG) of CNRS/INSU with INP and IN2P3, co-funded by CEA and CNES. 

%\bibliography{/Users/martin/Work/Papers/Biblio}
%\bibliographystyle{apj}

% Bibtex will create a .bbs file in the directory and before sending to the editor, I should replace the bibliography call by this file.

\end{document}